# TOWARDS TRUSTED VOLUNTEER GRID ENVIRONMENTS

Maher Khemakhem[1] and Abdelfettah Belghith[2]

[1]Department of Computer Science, ISG, Sousse University, Tunisia
`maher.khemakhem@fsegs.rnu.tn`
[2]Department of Computer Science, ENSI, Manouba University, Tunisia
`abdelfattah.belghith@ensi.rnu.tn`

## ABSTRACT

*Intensive experiences show and confirm that grid environments can be considered as the most promising way to solve several kinds of problems relating either to cooperative work especially where involved collaborators are dispersed geographically or to some very greedy applications which require enough power of computing or/and storage. Such environments can be classified into two categories; first, dedicated grids where the federated computers are solely devoted to a specific work through its end. Second, Volunteer grids where federated computers are not completely devoted to a specific work but instead they can be randomly and intermittently used, at the same time, for any other purpose or they can be connected or disconnected at will by their owners without any prior notification. Each category of grids includes surely several advantages and disadvantages; nevertheless, we think that volunteer grids are very promising and more convenient especially to build a general multipurpose distributed scalable environment. Unfortunately, the big challenge of such environments is, however, security and trust. Indeed, owing to the fact that every federated computer in such an environment can randomly be used at the same time by several users or can be disconnected suddenly, several security problems will automatically arise. In this paper, we propose a novel solution based on identity federation, agent technology and the dynamic enforcement of access control policies that lead to the design and implementation of trusted volunteer grid environments.*

## 1. INTRODUCTION

Conducting an online cooperative work trough the internet between collaborators who are dispersed geographically was a dream which, fortunately, has become a reality thanks to new technologies of information and telecommunication [14]. In the same manner, to provide enough power of computing or/and storage to some specific greedy applications to speedup the corresponding execution time according to the users needs was also a dream which, fortunately, has also become a tacit reality.

Possibilities of providing virtual organizations (as virtual huge computers) that allow, both, the cooperative work between collaborators who are dispersed geographically and/or the required speedup of any given greedy applications are, hence, a real fact [14]. These huge virtual computers are rather the result of federating a very huge number of computers belonging to several organizations. This phenomenon is commonly known as grid environments.

We can distinguish, at least, two categories of grid environments: dedicated grids and volunteer grids. A dedicated grid environment is the result of the federation over the internet of a fixed and large number of devoted resources (computers) to some specific purposes. The utilization of such environment for any given work or purpose requires first its prior reservation from the corresponding managers. Mostly, users of this kind of grids have to prepare themselves the required methods that allow the optimal utilization of such infrastructures. Such infrastructures present at the same time several advantages and disadvantages. Security, fault tolerance and maybe an easy management are considered amongst the most attractive sides of such environments. However, the optimal resources' utilization which is due to the condemnation of a big number of computers and the corresponding cost are considered amongst the darkest sides of such environments.

A volunteer grid environment is, rather, the result of the benevolent federation over the internet of a dynamic varying number of non devoted resources (computers) provided by volunteer donators. Amongst the most attractive features of these infrastructures is flexibility which is demonstrated by the dynamically varying number of volunteer donators (i.e.; participating computers). Moreover, these federated computers can be exploited and used simultaneously by their local owners and some other distributed tasks launched by distant authorized users. Some conducted experiments show and confirm, in particular, that the computing power of a given grid computing can never be used 100% by a given distributed processing work [5]. So far, volunteer grids such as: [1], [2], [3], [4], [6], [7], are unable to guarantee the necessary trust and security to their users. This is considered amongst the most serious issues in volunteer grid computing environments.

Identity federation was successfully used to provide security for some given shared resources between a fixed number of dispersed organizations [8]. To the best of our knowledge, it has never been used in the case where the shared resources are dynamically varying in numbers and/or locations.

On the other hand, agent technology provides very interesting concepts such autonomy, mobility and intelligence which can be exploited to adapt identity federation for the case of volunteer grid environments where resources are dynamic. To better ensure the security and trust of these benevolent dynamic environments, we further suggest the utilization of the enforcement of the access control policies which is very convenient for mobile agents' security, [31], [32].

Consequently, this paper starts with the presentation of an overview on grid environments followed by another one on identity federation. Agent technology, the corresponding added values and the enforcement of their access control policies are presented in the fourth part. The proposed approach which attempt to provide the required trust and security for a given volunteer grid is detailed in the fifth part. Finally, we conclude and present some perspectives of this work.

## 2. GRID ENVIRONMENTS

A grid environment, commonly known as a grid computing, is a virtual organization obtained by the interconnection (federation) through the internet of many geographically dispersed resources and which provides several possibilities such as remote collaboration and huge power of computing and/or storage for a given purpose (or multiple purposes) [14], [26], [27], [28], [30]. Such environments are, consequently, very attractive because the corresponding users can share many heterogeneous computer resources such as files, software, computing power and data storage belonging to several organizations or users. A grid is a collection of machines, sometimes referred to as "nodes," "resources," "members," "donors," "clients," "hosts," "engines," and many other such terms. They all contribute, in any combination of resources, to the grid as a whole. Some resources may be used by all users of the grid while others may have specific restrictions. In most organizations, there are large amounts of under utilized computing resources. Most desktop machines are known to be busy for less than 5 percent of the time. In some organizations, even the server machines can often be relatively idle. Grid computing provides then a framework for exploiting these under utilized resources and thus has the possibility of substantially increasing the efficiency of resource usage. Often, machines may have enormous unused disk drive capacities. Grid computing, more specifically, a "data grid", can be used to aggregate this unused storage into a much larger virtual data store, possibly configured to achieve improved performance and reliability over that of any single machine. Consequently, a grid computing is an infrastructure that allows many institutions (regardless of their geographical locations) to interconnect a large collection of their heterogeneous computer and systems to share together a set of software and/or hardware resources, services, licenses, etc... This ability of sharing resources in various combinations will lead to many advantages such as:
- increase the efficiency of resource usage;
- ease the remote collaboration between: institutions, researchers, etc;
- provide users with a huge computing power;
- provide users with a huge storage capacity, etc.

## 2.1 Dedicated vs. Volunteer grids

Dedicated grids are commonly composed of a fixed and large number of federated computers [27], [28], [30] which can be used only by a fixed number of authorized users for some very specific purposes. These kinds of grids are then inherently well secured and trusted by the corresponding users. However, volunteer grids are composed of several connected and federated computers belonging to volunteer donators in a benevolent way (users, institutions, etc.), [2]. They aim at providing very large scale storage and computing power to their users (subscribers) for general purpose usage. Volunteer donators have just to subscribe their computers to be shared through some specific web sites without any engagement from their side especially in terms of trust and security.
As consequence, this kind of very interesting grid suffers of trust and security issues [2]. Hopefully, identity federation with agents' technology can be used to solve this problem as will be explained and elaborated next.

## 3. IDENTITY FEDERATION (AN OVERVIEW)

Identity federation aims at providing trust among a set of users (web services' consumers) and a set of web service providers. The key of this concept consists in implementing two basic components which are the identity provider (IP) and the service provider (SP) such that any user (consumer) has to login first to the identity provider to acquire an identity (as a key) which will allow him to consume the requested service from the adequate provider [8]. So, the acquired identity is a kind of a certificate to access to the provider side. Of course, communications between all involved parts is achieved in encrypted manners according to the well known standards and solutions such as SAML (Secure Assertion Markup Language), RBAC (Role Based Access Control) of OASIS, [29].

## 4. THE AGENT PARADIGM

The Agent paradigm is very attractive because it mimics human or animal societies or communities in terms of interactions and coordination to achieve together common goals or to solve complex problems [9], [14], [18], [17], [24]. An agent is a physical or logical entity that owns certain characteristics, some times looking like human ones. An agent is a semiautonomous entity like a human or an animal. It can conduct some tasks to achieve a goal, possibly without a central control and sometimes without a prior planning. In addition, an agent can interact and cooperate with other agents to achieve complex tasks. It is able sometimes to adapt its process face to some situations, e.g., environment changes. Overall, an agent can be intelligent (cognitive) or reactive, mobile or stationary [11], [15], [16], [19], [26].

## 4.1 Cognitive vs. reactive agents

An intelligent agent, also called cognitive, can act autonomously. It is able to achieve complex tasks without any help [10], [13], [14]. It owns a knowledge base that can be used to manage all its processes, including its skills and interactions with other agents. In addition, to achieve its goals, a cognitive agent can also make all necessary decisions, for example, to optimize the achievement of a task. The great advantage of such an agent is the simplicity with which it can be reused for many applications (or within different societies) because of its inherent modularity. However one main disadvantage of cognitive agents is the difficulty of their design, implementation and integration within an agent society or community. On the other hand, a reactive agent is not intelligent. It is considered as a part of its community. It acts only after receiving stimulation from its environment. In addition, it does not own specific goals and is unable to make decisions. Thus, such an agent is unable to solve tasks or achieve goals without the help of other agents. However, one main advantage of reactive agents is the relatively low complexity of their design, implementation and integration within an agent society.

### 4.2 Mobile vs. stationary agents

An agent can be either stationary if it is usually executed at the same location (for example on the same PC on a network), or mobile if it may be executed at different locations. An agent can be transferred to different locations for several reasons: special resource availability, execution load balancing, communication reduction, better QoS, etc. In fact, the use of mobile agents in some distributed applications is considered as a fundamental technology in next generation computing [20].

### 4.3 Multi agent systems

Agents are often grouped to solve complex, often distributed tasks. The grouped agents form a multi agent system where they communicate to achieve a common goal. A multi agent system mimics humans or animals societies in several aspects especially in terms of interaction, cooperation and sometimes negotiation to solve complex problems. A multi agent system can have the following advantages [15]: speedup due to concurrent processing, less communication bandwidth requirements because processing is located nearer the source of information, more reliability because of the lack of a single point of failure, improved responsiveness due to processing, sensing and effecting being collocated and finally an easier system development due to the modularity inherent to the decomposition into semiautonomous agents. In addition, several researchers believe that the agent paradigm is the best way to solve efficiently distributed problems. Within any multi agent system, communication is vital because agents solve collectively a given complex task by allocating one or more sub problem per agent. Thus, agent interaction and cooperation are considered among the major issues in multi agent system design and development. The difficulty resides in how to get agents to cooperate effectively.
Interaction is further complicated whenever more than one agent can solve the same sub problems or overlapped sub problems but with different algorithms or data. Many works have been proposed to solve interaction and cooperation problems [9], [15].

### 4.4 Agent security and enforcement of the access control policies

Secure communication between agents is considered as a challenge because of the inherent complexity. In fact, security of agents can be viewed at different levels such as agent authentication and message authentication. Consequently, several works have been proposed to solve such problems [21], [22], [23]. Unfortunately, agent mobility is also considered as another source of security threats e.g., when a mobile agent move from a node to another one. Hopefully, the enforcement of the access control policies is considered a very promising solution which can provide security to mobile agents, [31], [32], [36]. We distinguish four classes of such mechanisms:
    1. Static analysis;

    2. Execution monitoring;
    3. Program rewriting and;
    4. Execution monitoring combined with program rewriting.

### 4.4.1 Static analysis

Such a mechanism is based on a static analysis of any untrusted program before its execution. This analysis is always achieved in a finite time. Only accepted programs are able to be executed on the host node. Static type-checker for type safe languages of Java virtual machines, JFlow and standard virus scanners constitute good samples of static analysis environments. In fact, this mechanism had shown good performances for some specific applications. However, programmers, in this case, must produce manually, proofs of correction. Besides, these proofs are generally tedious and involved [31], [32], [33], [34], [35].

### 4.4.2 Execution monitoring

Such mechanism is based on the permanent control of any untrusted program during its execution over the host node. It can intercept the pertinent events related to the system security and intervenes whenever a violation of the security policies is detected. Firewalls, virtual machines and OS behave as execution monitors [31].

### 4.4.3 Program rewriting

Such mechanism is based on the program rewriting before its execution. It means that whenever an untrusted program arrives to the hosted node, the program rewriting mechanism try to modify the code of this program to be a trusted one according to the access control policies of this hosted node. This will facilitate its execution in a safe manner, [31], [33].

### 4.4.4 Execution monitoring combined with program rewriting

This mechanism was the latest one that we have encountered during our bibliographical study [31]. It consists of combining both execution monitoring and program rewriting using the aspect oriented programming. One of the advantages of the aspect oriented programming is its ability to separate the functional process (business processes) concerns from their technicalities (execution processes) [31], this will facilitate, certainly, the achievement of the targeted objective.

## 5. THE PROPOSED APPROACH

Identity federation is so far used to create a trust circle (environment) between a known set of providers and a known set of consumers (clients) of Web Services (WS) [8]. Our purpose here is, however, to create a trust environment between a variable set of volunteer computers and a variable set of authorized users of a given volunteer grid. By analogy to WS, we can view any connected computer of the grid as a provider which has been subscribed, first, by its donator as we have seen previously. However, this provider can be disconnected at any time without any prior notification. In the same manner, any authorized user of the grid can be seen as a service consumer. Moreover, the donator which is the source of the provided resource (provider) is, commonly, at the same time the user (consumer). Any utilization of any computer of the grid whether for a distributed storage or a distributed processing is considered as a WS. Consequently, the similarity between our problem and the WS is clear and evident. However, for the case of volunteer grids, providers (computers) can disappear (get disconnected) suddenly or intermittently by their donators (owners) at will. This situation makes our problem a little bit more complicated because we have to guarantee, at all time, to all users of the grid trust and security. This means that no perturbation of their engaged work on the grid is allowed to occur. Agent technology provides autonomy, intelligence and mobility which can be exploited to solve this problem. Unfortunately, mobile agents are considered also as a source of some security problems. Fortunately, the dynamic enforcement of access control policies can ensure the security of mobiles agents and lead consequently to an effective solution to our problem. The basic idea of our proposal consists to implement the two required components for the identity federation which are, in order, the identity provider and the service provider within the same multi agent system (MAS). Where the identity provider (IP) component serves to:
- Subscribe any provider/user according to the international security standards and solutions W3C, [25], OASIS, [29];
- Disseminate the security rules and dynamic policies especially for the authentication process among the connected computers (providers) whenever the situation changes (i.e. a new provider is connected or an existing provider is disconnected).

And where the service provider (SP) component serves to:
- Authenticate any user with the help of the IP;
- Provide the storage or processing service to any authorized user.

As we have seen, the IP and SP components will be implemented within the same MAS which will be composed, consequently, of one Stationary agent and a set of Mobile agents.

## 5.1 The MAS architecture

In fact, the stationary agent of the proposed MAS plays the role of the coordinator among the mobiles one and achieve the IP component mission. It includes in addition to communication capability the following skills:
- Specific methods to authenticate and interface known users. Of course, we suppose that this agent owns a extensible knowledge base and the adequate tools that allow it to authenticate and keep track of every transaction with any user by using the dynamic enforcement of access control policies;
- The ability to subscribe any new trusted user;
- The ability of negotiation with the remaining agents;
- The ability to instantiate a mobile agent whenever a provider or user is subscribed. This mobile agent is, in fact, the provided identity (certificate or passport) to the user that will allow him to use the grid (as provider or user);
- The ability to disseminate any news among the remaining agents to update their knowledge base;
- The ability to destroy any mobile agent whenever the corresponding provider or user is disconnected;
- The ability to make useful decisions such as which computers will participate to any given work;
- The ability to protect itself against hackers and untrusted environments. This agent is supposed to be activated 24 hours on 24 hours 7 days on 7 days and implemented over a specific trusted node which will form the kernel of the grid.

Consequently, the proposed MAS will be implemented, at the beginning, as two agents, the stationary one which is already presented and a mobile one which will be instantiated whenever a provider (computer) or a user is connected (subscribed).

This mobile agent includes in addition to communication capability the following skills:
- The ability of instantiation;
- The ability of negotiation;
- The ability of migration through the grid network from the IP node (fixed) to any authorized SP one (volunteer);
- The ability to authenticate every authorized SP node by using the dynamic enforcement of access control policies;
- The ability to make some decisions such as informing the coordinator that the corresponding provider is unable to achieve the requested work;

- The ability to protect itself against hackers untrusted environments for example by destroying itself according to the dynamic enforcement of the access control policies.

### 5.2 The MAS mechanism

As we have seen, we suppose that any user of such volunteer grid is, at the same time, a resource donator. So, whenever a new user (donator) wants to participate to the sharing and utilization of this grid, he must first request for a subscription through the IP (stationary agent) interface. The IP can accept or reject this subscription depending on the satisfaction of some predefined security rules, policies and constraints which should conform to the international security standards and solutions W3C [25], OASIS [29]. Suppose that a given new subscriber is accepted, so if he is connected, then a corresponding mobile agent is instantiated to allow him the sharing of the grid (as an access granted). This mobile agent, which plays the role of an access granted, can be instantiated whenever the need appear such as for example when the authorized user would like to communicate with another authorized user or to submit a job to a specific provider. The instantiation process is achieved according to a hierarchical model. The first instance (called also the main instance) will be destroyed whenever the user is disconnected. The longevity of any other instance (called also a secondary instance) is limited by the duration of the corresponding mission in charge. Of course, each new instance can inherit all or a part of the functionality of its parent according to the mission needs. It means that if an instance is created to handle a given task execution on a specific provider then, it will inherit only the required functionality from its parent to achieve this mission and so on. If an instantiated mobile agent is intercepted by a hacker, then it proceeds to destroy itself after sending a notification to its parent. In the same manner, if a mobile agent encounters any problem whatever the location and the situation, it proceeds to destroy itself after sending a notification to its parent. After receiving this notification, the parent agent has to take a decision about the target (provider) of the new instance that will achieve the uncompleted work at hand until its achievement.

### 6. CONCLUSION AND PERSPECTIVES

In this paper, we focused on the trust and security issues in volunteer grid environments. Despite their security problematic issues, such benevolent environments remain very promising for they present several advantages such as flexibility, resource availability and efficiency of use.
Our proposed model is based on identity federation, agent technology and the dynamic enforcement of access control policies and it constitutes an attempt to provide adequate level of trust and security for volunteer grid environments. Besides, our trust and security model can be easily integrated within existing volunteer grids based on international security standards and solutions. Further investigations are currently under way. In particular, experiments are being conducted to ascertain the viability and the efficiency of the proposed security and trust model.